\documentclass[journal]{IEEEtran}
\usepackage{amssymb}
\usepackage{amsfonts}
\usepackage{booktabs}
\usepackage[pdftex]{graphicx}
\usepackage[cmex10]{amsmath}
\usepackage{algorithm}
\usepackage{algorithmic}
\usepackage{array}
\usepackage[caption=false,font=footnotesize]{subfig}
\usepackage{colortbl}

\usepackage{verbatim}

 \allowdisplaybreaks[4]
\newcounter{MYtempeqncnt}

\begin{document}
\title{On the Achievability of Interference Alignment for Three-Cell Constant Cellular Interfering Networks}
\author{Yanjun~Ma,
        Jiandong~Li,~\IEEEmembership{Senior~Member,~IEEE},
        Rui~Chen,~\IEEEmembership{Student~Member,~IEEE},  and Qin~Liu
\thanks{This work was supported in part by the National Science Fund for Distinguished
Young Scholars under Grant 60725105, by the National Basic Research Program of China
 under Grant 2009CB320404, by the Program for Changjiang Scholars and Innovative
 Research Team in University under Grant IRT0852, by the Key Project of Chinese Ministry of Education under Grant
  107103, and by the 111 Project under Grant B08038.}
\thanks{The authors are with the State Key Laboratory of Integrated
Service Networks, Xidian University, Xi'an 710071, China
 (e-mail: \{yjm, jdli, qinliu\}@mail.xidian.edu.cn, rchenxidian@gmail.com).}}
\maketitle

\begin{abstract}
For a three-cell constant cellular interfering network, a new property of
alignment is identified, i.e., interference alignment (IA) solution obtained in
an user-cooperation scenario can also be applied in a non-cooperation
environment. By using this property, an algorithm is proposed by jointly
designing transmit and receive beamforming matrices. Analysis and numerical
results show that more degree of freedom (DoF) can be achieved compared with
conventional schemes in most cases.
\end{abstract}

%\IEEEpeerreviewmaketitle

%\newpage

\section{Introduction}
\IEEEPARstart{I}{nterference} alignment (IA) is a promising technique to
mitigate interference in wireless communication systems. It was shown that
$\frac{K}{2}$ DoF is achievable per time, frequency or antenna dimension in a
$K$-user interference channel (IC) \cite{ref1}. For a $K$-user constant MIMO
IC, IA based schemes were introduced in \cite{ref2}-\cite{ref5}, where it was
shown that more DoF is achievable than that of conventional schemes. For a
constant cellular interfering network, it was shown in \cite{ref6} that their
scheme provides respectable gain for a 19 hexagonal wrap-around-cell layout.
However, interference-free DoF is only achievable for a two-cell layout. It was
shown in \cite{ref7} that optimal DoF is achievable when $\lceil \frac{3}{2}N
\rceil \leq M < 2N$ for a two-cell MIMO interfering broadcast channel, where
each transmitter is equipped with $M$ antennas and each receiver is equipped
with $N$ antennas.

In this letter, we focus on a three-cell constant cellular interfering network
by using a new property of alignment, i.e., \emph{IA solution obtained in an
user-cooperation scenario can also be applied in a non-cooperation
environment}. We assume that each base station (BS) is equipped with $M$
antennas and each mobile station (MS) is equipped with $N$ antennas, and $M >
N$ which is most possible in a practical environment. We also assume there are
$K$ cell-edge users per cell where $K>1$, and each user sends $d$ streams to
its served BS simultaneously. We show that totally $3Kd$ DoF is achievable if
$M =KN$ and $d \leq \lfloor \frac{M}{3K-1} \rfloor$ or if $M < KN $ and $d \leq
\min \big\{\lfloor \frac{M}{3K-1} \rfloor, 3(KN-M) \big\}$. Numerical results
show that more DoF can be achieved compared with conventional schemes in most
cases.

\section{System Model}
For a three-cell constant cellular interfering network (an example is shown in
Fig.~\ref{fig1}), we assume that each BS is equipped with $M$ antennas, each MS
is equipped with $N$ antennas, and there are $K$ cell-edge users per cell. For
notation convenience, we refer to the $j$-th user in the $i$-th cell as user
$[i,j]$. For ease of analysis, we consider an uplink scenario\footnote{By using
the reciprocity of alignment \cite{ref5}, our scheme can also be applied in a
downlink scenario.}, and assume that each user tries to convey $d$ data streams
to its served BS by using a normalized precoding matrix $\textbf{W}^{[ij]}$.
Then we have
\begin{equation}
\mathbf{x}^{[ij]} = \textbf{W}^{[ij]} \textbf{s}^{[ij]},
\end{equation}
where $\textbf{s}^{[ij]}$ is a $d \times 1$ vector, which denotes the
transmitted data streams from user $[i,j]$, and satisfies an average power
constraint, i.e., $\mathbb{E} \left[\| \textbf{s}^{[ij]} \|^2 \right] \leq P$.
The received signal at the $i$-th BS is represented as
\begin{equation}
 \textbf{y}^{[i]} = \sum^3_{k=1} \sum^K_{j=1} \textbf{H}^{[kj]}_{i}\textbf{W}^{[kj]} \textbf{s}^{[kj]}
 +\textbf{n}^{[i]},
\end{equation}
where $\textbf{n}^{[i]} \sim \mathcal{N}(\textbf{0}, \sigma^2\textbf{I})$ is
the $M \times 1 $ additive white Gaussian noise, and $\textbf{H}^{[kj]}_{i}$ is
the $M \times N$ channel matrix from user $[k, j]$ to the $i$-th BS. The
channel is assumed to be constant over time, and perfect channel state
information (CSI) is available at all BSs and MSs. The $i$-th BS decodes the
desired signal for user $[i,j]$ by multiplying the cascaded receive beamforming
matrices, $\textbf{V}^{[i]}$ and $\textbf{P}_j^{[i]}$, and we obtain the
desired signal for user $[i,j]$
\begin{equation}
 \widetilde{\textbf{y}}^{[ij]} =\textbf{P}_j^{[i]\dag} \textbf{V}^{[i]\dag}  \sum^3_{k=1} \sum^K_{j=1} \textbf{H}^{[kj]}_{i}\textbf{W}^{[kj]} \textbf{s}^{[kj]}
 +\widetilde{\textbf{n}}^{[ij]},
\end{equation}
where $\textbf{V}^{[i]}$ is the normalized inter-cell interference (ICI)
elimination matrix, $\textbf{P}_j^{[i]}$ is the normalized inter-user
interference (IUI) elimination matrix, and $\widetilde{\textbf{n}}^{[ij]}
=\textbf{P}_j^{[i]\dag} \textbf{V}^{[i]\dag} \textbf{n}^{[i]}$ is the effective
noise vector. The notation $(\cdot)^{\dag}$ stands for conjugate transpose. Let
$\textbf{P}^{[i]} = \{\textbf{P}_1^{[i]}, \dots, \textbf{P}_K^{[i]}\}$ denote
the combined IUI elimination matrix at the $i$-th BS. We define the DoF region
as the following \cite{ref1}:
\begin{multline}
\mathcal{D} = \bigg\{ (d^{[11]}, \dots, d^{[3K]}) \in \mathbb{R}^{3K}_+ |
\forall
(\omega_{11}, \dots, \omega_{3K} ) \in \mathbb{R}^{3K}_+, \\
\sum^3_{i=1} \sum^K_{j=1} \omega_{ij} d^{[ij]} \leq \limsup_{\textrm{SNR}
\rightarrow \infty} \Big[ \sup_{\textbf{R} \in \mathcal{C}} \frac{1}{\log
\textrm{SNR}} \sum^3_{i=1} \sum^K_{j=1} \omega_{ij} R^{[ij]} \Big]\bigg\},
\end{multline}
where $\mathcal{D}$ is the capacity region, $\textrm{SNR} = P/ \sigma^2$, and
$R^{[ij]}$ is the rate of user $[i, j]$. Let
\begin{equation}
\eta  = \sum_{i \in \{ 1, 2, 3\}} \sum_{j \in \{ 1, \dots, K \}} d^{[ij]}
\end{equation}
be the total DoF in the network.

\section{An IA Based Scheme for Three-Cell Constant Cellular Interfering Networks}
\begin{figure}[!tr]
\centering
\includegraphics[width=7.5cm]{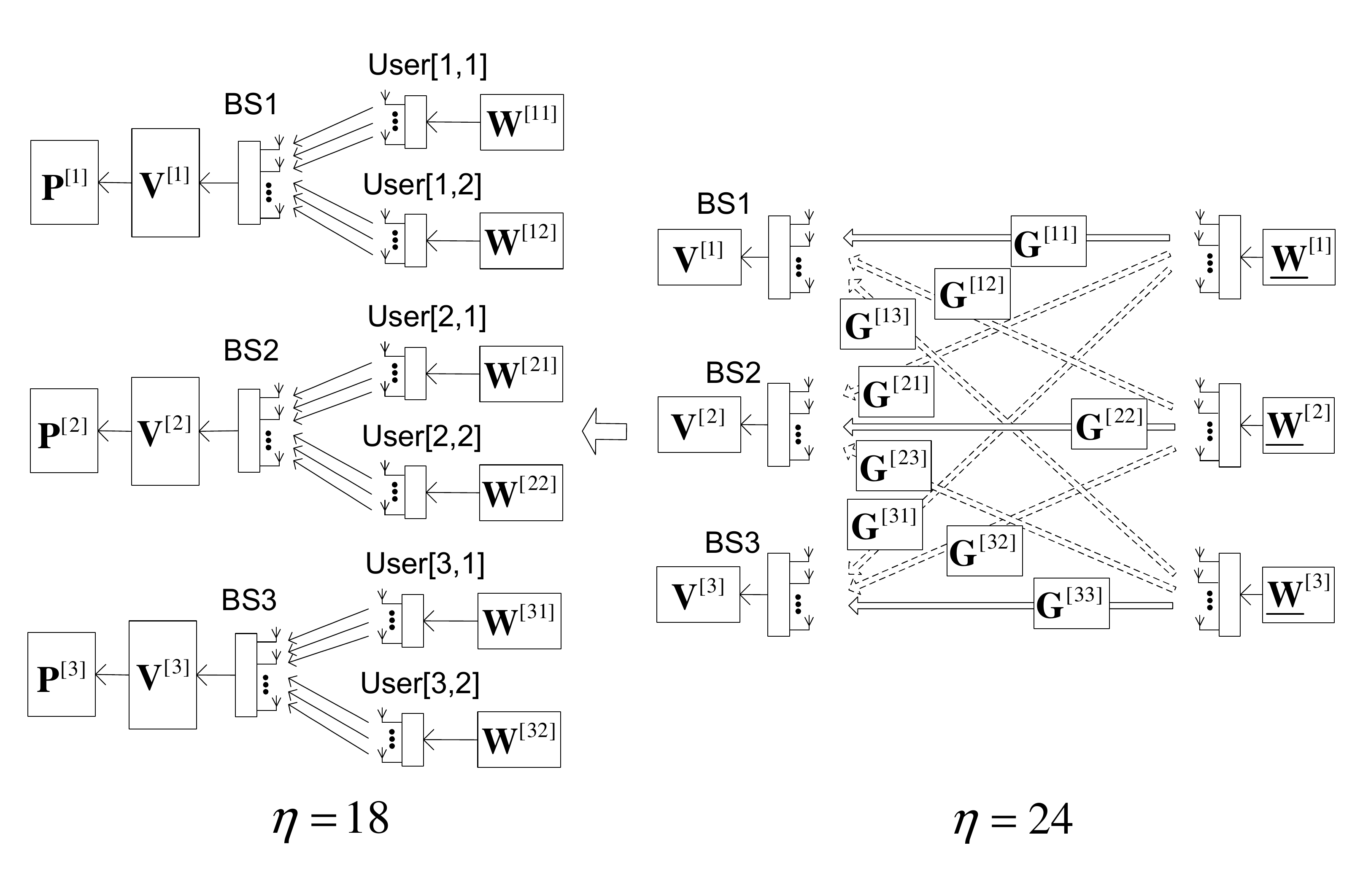}
\caption{An illustration for a three-cell constant cellular interfering
network, where $M=16$, $N=8$, $K=2$, and $d=3$.} \label{fig1}
\end{figure}

In this section, an IA based scheme is introduced for the three-cell constant
cellular interfering network. A motivating example is given first (as is shown
in Fig.~\ref{fig1}), where $M=16$, $N=8$, $K=2$, and $d=3$. We show that
totally 18 DoF is achievable in this scenario.

We divide our scheme into two phases. First, ICI is aligned into a smaller
vector space at each BS by joint design of all the precoding matrices. Second,
IUI is eliminated through cascaded receive beamforming matrices at each BS.

\textbf{Phase I: ICI alignment.} By applying the IA solution obtained in an
MIMO IC to a cellular environment, which is presented in Fig. \ref{fig1}, we
show that 12 ICI streams at each BS can be aligned into a vector space of 9
dimensions simultaneously, i.e.,
\begin{multline}\label{eq4}
\text{dim}\Big\{ \text{span}\big( [ \textbf{H}_1^{[21]} \textbf{W}^{[21]} ~
\textbf{H}_1^{[22]} \textbf{W}^{[22]} ~\\
 \textbf{H}_1^{[31]} \textbf{W}^{[31]} ~
\textbf{H}_1^{[32]} \textbf{W}^{[32]} ] \big) \Big\}= 9,
\end{multline}
\begin{multline}\label{eq5}
\text{dim}\Big\{\text{span}\big( \big[ \textbf{H}_2^{[11]} \textbf{W}^{[11]}~
\textbf{H}_2^{[12]} \textbf{W}^{[12]} ~\\ \textbf{H}_2^{[31]} \textbf{W}^{[31]}
~ \textbf{H}_2^{[32]} \textbf{W}^{[32]} \big] \big) \Big\} = 9,
\end{multline}
\begin{multline}\label{eq6}
\text{dim}\Big\{\text{span}\big( \big[ \textbf{H}_3^{[11]} \textbf{W}^{[11]} ~
\textbf{H}_3^{[12]} \textbf{W}^{[12]} ~\\ \textbf{H}_3^{[21]} \textbf{W}^{[21]}
~ \textbf{H}_3^{[22]} \textbf{W}^{[22]} \big] \big) \Big\}= 9.
\end{multline}

Let $\underline{\textbf{W}}^{[i]} = \big[\textbf{W}^{[i1]\dag}~
\textbf{W}^{[i2]\dag} \big]^{\dag}$ be the combined transmit precoding matrix
of all cell-edge users in the $i$-th cell. Let $\textbf{G}^{[ij]} =\big[
\textbf{H}_i^{[j1]}~ \textbf{H}_i^{[j2]} \big]$ represent the combined channel
matrix. Then the effective channel is a three-user MIMO IC where each node is
equipped with $M=16$ antennas. Following the analysis in \cite{ref1}, there
exists a $16\times 8$
 $\underline{\textbf{W}}^{[i]}$, $i=\{1,2,3\}$, satisfies (\ref{eq123}) -
(\ref{eq125}).
\begin{equation}\label{eq123}
\text{span} \left[ \textbf{G}^{[12]} \underline{\textbf{W}}^{[2]} \right] =
\text{span} \left[    \textbf{G}^{[13]} \underline{\textbf{W}}^{[3]} \right],
\end{equation}
\begin{equation}\label{eq124}
\text{span}\left[  \textbf{G}^{[21]} \underline{\textbf{W}}^{[1]} \right] =
\text{span} \left[   \textbf{G}^{[23]} \underline{\textbf{W}}^{[3]} \right],
\end{equation}
\begin{equation}\label{eq125}
\text{span} \left[   \textbf{G}^{[31]} \underline{\textbf{W}}^{[1]} \right] =
\text{span} \left[  \textbf{G}^{[32]} \underline{\textbf{W}}^{[2]} \right].
\end{equation}
Let
\begin{equation}
\textbf{E} = ( \textbf{G}^{[31]} )^{-1} \textbf{G}^{[32]}  ( \textbf{G}^{[12]}
)^{-1} \textbf{G}^{[13]} ( \textbf{G}^{[23]} )^{-1} \textbf{G}^{[21]},
\end{equation}
\begin{equation}
\textbf{F} = ( \textbf{G}^{[32]} )^{-1} \textbf{G}^{[31]},
\end{equation}
\begin{equation}
\textbf{C} = ( \textbf{G}^{[23]} )^{-1} \textbf{G}^{[21]}.
\end{equation}
Then
\begin{equation}
\text{span}(  \underline{\textbf{W}}^{[1]}   ) = \text{span}(  \textbf{E}
\underline{\textbf{W}}^{[1]} ),
\end{equation}
\begin{equation}\label{eq121}
\underline{\textbf{W}}^{[2]}    =  \textbf{F} \underline{\textbf{W}}^{[1]},
\end{equation}
\begin{equation}\label{eq122}
\underline{\textbf{W}}^{[3]}     =  \textbf{C} \underline{\textbf{W}}^{[1]} .
\end{equation}

\begin{figure}[!tr]
\centering
\includegraphics[width=7.5cm]{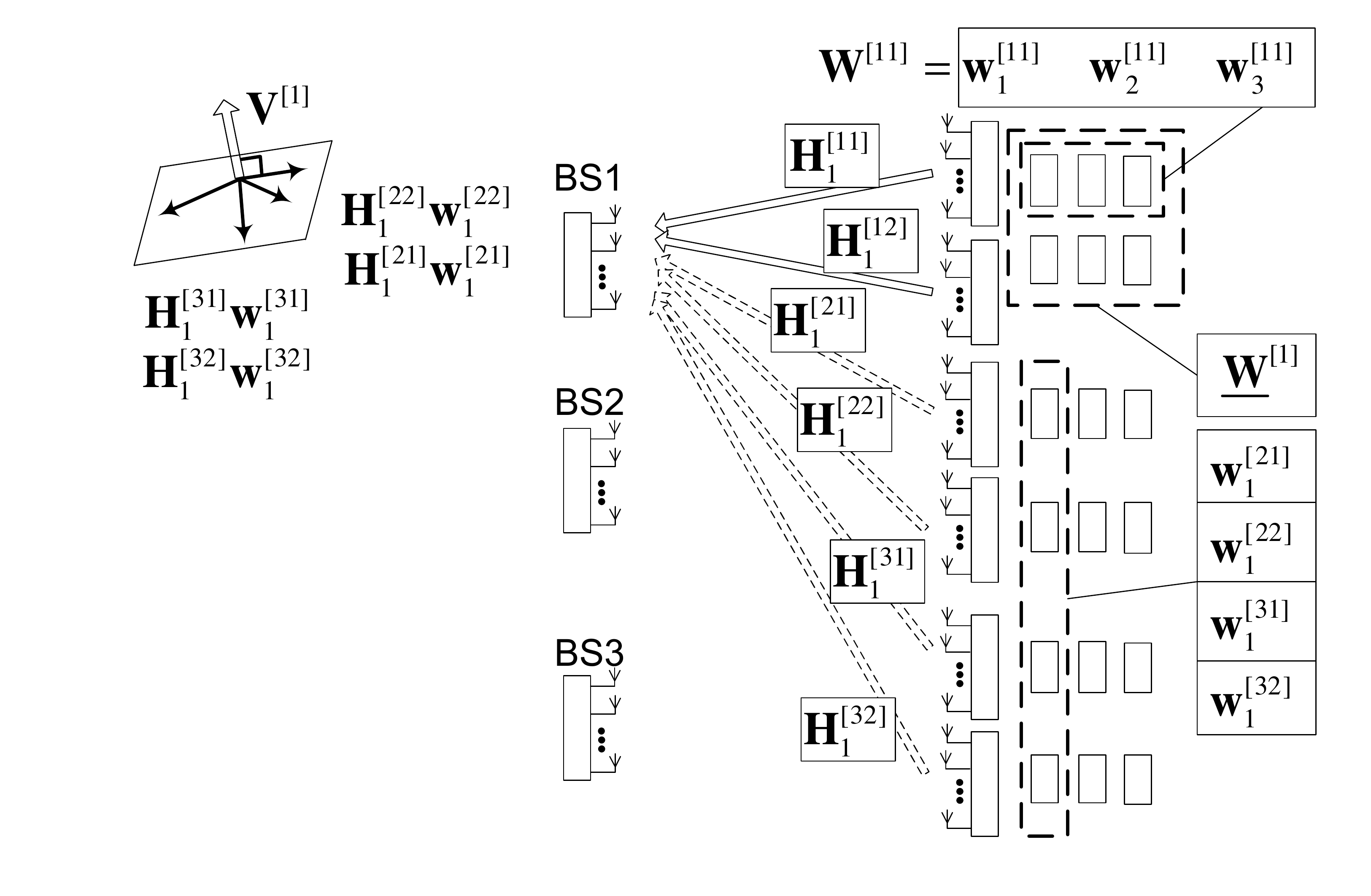}
\caption{In this illustration, it is shown that 4 interference signals are
aligned into a vector space of 3 dimensions.} \label{fig22}
\end{figure}

So, if we allow full user-cooperation, 24 DoF is achievable in this scenario.
However, if user-cooperation is not allowed, we show that (\ref{eq4}) -
(\ref{eq6}) are also satisfied in the following.

Let $\{\textbf{e}_1, \dots, \textbf{e}_{16}\}$ be the eigenvectors of
$\textbf{E}$. Let $\underline{\textbf{W}}^{[1]}$ be any three eigenvectors of
$\textbf{E}$, for example, let $\underline{\textbf{W}}^{[1]} =
\{\textbf{e}_1,\textbf{e}_2, \textbf{e}_3\}$, and calculate
$\underline{\textbf{W}}^{[2]}$ and $\underline{\textbf{W}}^{[3]}$ according to
(\ref{eq121}) and (\ref{eq122}), respectively.
 Let $\textbf{W}^{[ij]} = [
\textbf{w}_1^{[ij]}~\textbf{w}_2^{[ij]}~\textbf{w}_3^{[ij]}  ]$, $i \in
\{1,2,3\}$, $j \in \{1,2\}$ (Fig.~\ref{fig22} shows the relationship among
$\underline{\textbf{W}}^{[i]}$, $\textbf{W}^{[ij]}$, and
$\textbf{w}_k^{[ij]}$). We rewrite (\ref{eq123}) as
\begin{equation*}\label{eq432}
\text{span}\Big(
\begin{bmatrix}\textbf{H}_1^{[21]}
\textbf{H}_1^{[22]}
\end{bmatrix}
\begin{bmatrix}
\textbf{w}_k^{[21]}\\ \textbf{w}_k^{[22]}
\end{bmatrix}
\Big) = \text{span}\Big(
\begin{bmatrix}\textbf{H}_1^{[31]}
\textbf{H}_1^{[32]}
\end{bmatrix}
\begin{bmatrix}
\textbf{w}_k^{[31]}\\ \textbf{w}_k^{[32]}
\end{bmatrix}
 \Big),
\end{equation*}
where $k\in \{1,2,3\}$, i.e.,
\begin{equation*}
 \text{span} (  \textbf{H}_1^{[21]} \textbf{w}_k^{[21]} +
 \textbf{H}_1^{[22]} \textbf{w}_k^{[22]} ) = \text{span}(
  \textbf{H}_1^{[31]} \textbf{w}_k^{[31]}  +  \textbf{H}_1^{[32]} \textbf{w}_k^{[32]}
  ).
\end{equation*}
So, there exist non-zero $\alpha_{1k}$ and $\alpha_{2k}$ satisfy
\begin{equation*}\label{eq434}
 \alpha_{1k} (  \textbf{H}_1^{[21]} \textbf{w}_k^{[21]} +
 \textbf{H}_1^{[22]} \textbf{w}_k^{[22]} ) = \alpha_{2k}(
  \textbf{H}_1^{[31]} \textbf{w}_k^{[31]}  +  \textbf{H}_1^{[32]} \textbf{w}_k^{[32]}
  ),
\end{equation*}
i.e.,
\begin{multline}
\text{dim}\Big\{ \text{span}\big( \big[ \textbf{H}_1^{[21]} \textbf{w}_k^{[21]}
~ \textbf{H}_1^{[22]} \textbf{w}_k^{[22]}~\\
 \textbf{H}_1^{[31]} \textbf{w}_k^{[31]} ~
\textbf{H}_1^{[32]} \textbf{w}_k^{[32]} \big] \big) \Big\}=3.
\end{multline}
When the channels are generic, i.e., the elements of the channel matrices are
randomly and independently generated from continuous distributions, we have
\begin{multline}
\text{dim}\Big\{ \text{span}\big( \big[ \textbf{H}_1^{[21]} \textbf{W}^{[21]} ~
\textbf{H}_1^{[22]} \textbf{W}^{[22]} ~ \textbf{H}_1^{[31]} \textbf{W}^{[31]} ~
\textbf{H}_1^{[32]} \textbf{W}^{[32]} \big] \big) \Big\}\\
= \text{dim}\Big\{ \text{span}\big( \big[ \textbf{H}_1^{[21]}
\textbf{w}_1^{[21]} ~  \textbf{H}_1^{[22]} \textbf{w}_1^{[22]}~
 \textbf{H}_1^{[31]} \textbf{w}_1^{[31]} ~
\textbf{H}_1^{[32]} \textbf{w}_1^{[32]} \big] \big) \Big\}\\
+ \text{dim}\Big\{ \text{span}\big( \big[ \textbf{H}_1^{[21]}
\textbf{w}_2^{[21]} ~  \textbf{H}_1^{[22]} \textbf{w}_2^{[22]}~
 \textbf{H}_1^{[31]} \textbf{w}_2^{[31]} ~
\textbf{H}_1^{[32]} \textbf{w}_2^{[32]} \big] \big) \Big\}\\
+ \text{dim}\Big\{ \text{span}\big( \big[ \textbf{H}_1^{[21]}
\textbf{w}_3^{[21]} ~  \textbf{H}_1^{[22]} \textbf{w}_3^{[22]}~
 \textbf{H}_1^{[31]} \textbf{w}_3^{[31]} ~
\textbf{H}_1^{[32]} \textbf{w}_3^{[32]} \big] \big) \Big\}\\
=9.~~~~~~~~~~~~~~~~~~~~~~~~~~~~~~~~~~~~~~~~~~~~~~~~~~~~~~~~~~~~~~~~~~~~~
\end{multline}
All the interference at BS1 has been aligned into a vector space of 9
dimensions. Along the same way, (\ref{eq5}) and (\ref{eq6}) are also satisfied.

\begin{figure}[!tr]
\centering
\includegraphics[width=6.5cm]{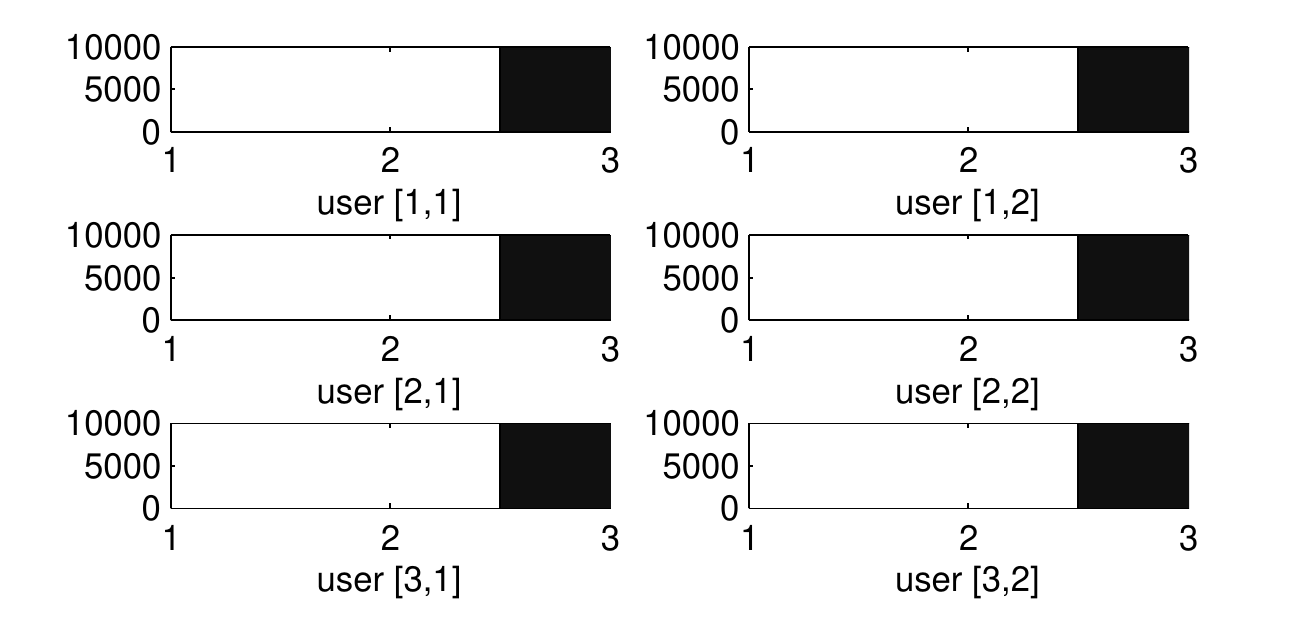}
\caption{Rank distribution of $\textbf{W}^{[ij]}$ at user $[i,j]$ for the
motivating example, where 10000 Monte Carlo tests are performed.} \label{fig55}
\end{figure}

\emph{Remark:} When the channels are generic, $\textbf{e}_k$ will be a random
vector, so are $\textbf{w}_k^{[i1]}$ and $\textbf{w}_k^{[i2]}$, where
$\textbf{e}_k=[\textbf{w}_k^{[i1]\dag}~\textbf{w}_k^{[i2]\dag}]^{\dag}$. Then
$\textbf{W}^{[ij]}=[\textbf{w}_1^{[ij]}~\textbf{w}_2^{[ij]}~\textbf{w}_3^{[ij]}
]$ will be full rank, i.e., $\textrm{rank}(\textbf{W}^{[ij]})=3$, with
probability 1. Fig. \ref{fig55} shows rank distribution of $\textbf{W}^{[ij]}$
at user $[i,j]$.

There are 7 interference-free dimensions left at each BS, and each BS can
decode 6 streams sent from its cell-edge users. Then we choose
\begin{equation*}
\textbf{V}^{[1]} \subseteq \text{null}\Big( \big[ \textbf{H}_1^{[21]}
\textbf{W}^{[21]} ~~  \textbf{H}_1^{[22]} \textbf{W}^{[22]} ~~
\textbf{H}_1^{[31]} \textbf{W}^{[31]} ~~ \textbf{H}_1^{[32]} \textbf{W}^{[32]}
\big] \Big),
\end{equation*}
\begin{equation*}
\textbf{V}^{[2]} \subseteq \text{null}\Big( \big[ \textbf{H}_2^{[11]}
\textbf{W}^{[11]}~~ \textbf{H}_2^{[12]} \textbf{W}^{[12]} ~~
\textbf{H}_2^{[31]} \textbf{W}^{[31]} ~~  \textbf{H}_2^{[32]} \textbf{W}^{[32]}
\big] \Big),
\end{equation*}
\begin{equation*}
\textbf{V}^{[3]} \subseteq \text{null}\Big( \big[ \textbf{H}_3^{[11]}
\textbf{W}^{[11]} ~~  \textbf{H}_3^{[12]} \textbf{W}^{[12]} ~~
\textbf{H}_3^{[21]} \textbf{W}^{[21]} ~~  \textbf{H}_3^{[22]} \textbf{W}^{[22]}
\big] \Big).
\end{equation*}

\textbf{Phase II: IUI elimination.} We finally obtain an ICI-free channel,
i.e., $\overline{\textbf{H}}_j^{[jk]} = \textbf{V}^{[j]\dag}
\textbf{H}_j^{[jk]} \textbf{W}^{[jk]}$, $j\in  \{1,2,3\}$ and $k\in \{1,2\}$.
The $j$-th BS calculates $\textbf{P}_1^{[j]} =
\text{null}(\overline{\textbf{H}}_j^{[j2]})$, $\textbf{P}_2^{[j]} =
\text{null}(\overline{\textbf{H}}_j^{[j1]})$, and IUI is eliminated.

Then 18 streams can be sent simultaneously without any interference. If
conventional schemes, such as orthogonal schemes are used, at most 16 DoF is
achievable. For general antenna configuration and general number of users per
cell in a three-cell multi-user MIMO environment when $M>N$, we have the
following theorem.
\newtheorem{theorem}{Theorem}
\begin{theorem}
In a three-cell constant interfering network, we assume that each BS is
equipped with $M$ antennas, each MS is equipped with $N$ antennas, $K$
cell-edge users are served simultaneously per cell, and each user sends $d$
streams to its served BS. If $M =KN$ and $d \leq \lfloor \frac{M}{3K-1}
\rfloor$ or if $M < KN $ and $d \leq \min \big\{\lfloor \frac{M}{3K-1} \rfloor,
3(KN-M) \big\}$, then $\eta=3Kd$ DoF is achievable.
\end{theorem}

\begin{figure*}[!t]
\normalsize
% Store the current equation number.
\setcounter{MYtempeqncnt}{\value{equation}}
% Set the equation number to one less than the one
% desired for the first equation here.
% The value here will have to changed if equations
% are added or removed prior to the place these
% equations are referenced in the main text.
\setcounter{equation}{31}
\begin{equation}\label{eq77}
 \overline{\textbf{H}}=
\begin{bmatrix}
&0 &\dots &0  &\textbf{H}_1^{[21]} &\dots &\textbf{H}_1^{[2K]} &\textbf{H}_1^{[31]} &\dots &\textbf{H}_1^{[3K]} \\
& \textbf{H}_2^{[11]}  & \dots   & \textbf{H}_2^{[1K]}  &0   &\dots  &0  & \textbf{H}_2^{[31]} & \dots & \textbf{H}_2^{[3K]}\\
& \textbf{H}_3^{[11]} & \dots    & \textbf{H}_3^{[1K]}  & \textbf{H}_3^{[21]} & \dots & \textbf{H}_3^{[2K]} &0 &\dots &0\\
\end{bmatrix}
\end{equation}
% Restore the current equation number.
\setcounter{equation}{21}
% IEEE uses as a separator
\hrulefill
% The spacer can be tweaked to stop underfull vboxes.
\vspace*{4pt}
\end{figure*}

\begin{proof}
When $M = KN$ and $d \leq \lfloor\frac{M}{3K-1}\rfloor$, we combine all
cell-edge users into an effective user. Let $\textbf{G}^{[ij]} =\left[
\textbf{H}_i^{[j1]}~ \dots~ \textbf{H}_i^{[jK]} \right]$ be the combined
channel matrix. Let $\underline{\textbf{W}}^{[i]} =
\left[\textbf{W}^{[i1]\dag}~ \dots~ \textbf{W}^{[iK]\dag} \right]^{\dag}$
represent the combined transmit precoding matrix. Then the effective channel is
a three-user IC where each node is equipped with $M$ antennas. Following the
analysis in \cite{ref1}, there exists a $KN \times d$
$\underline{\textbf{W}}^{[i]}$, $i=\{1,2,3\}$, satisfies (\ref{eq7111}) -
(\ref{eq9111}) as $d \leq \lfloor\frac{M}{3K-1}\rfloor < \frac{M}{2}$.
\begin{equation}\label{eq7111}
\text{span} (\textbf{G}^{[12]} \underline{\textbf{W}}^{[2]} ) = \text{span} (
 \textbf{G}^{[13]} \underline{\textbf{W}}^{[3]} ),
\end{equation}
\begin{equation}\label{eq8111}
\text{span}(  \textbf{G}^{[21]} \underline{\textbf{W}}^{[1]} ) = \text{span} (
\textbf{G}^{[23]} \underline{\textbf{W}}^{[3]} ),
\end{equation}
\begin{equation}\label{eq9111}
\text{span}  (  \textbf{G}^{[31]}  \underline{\textbf{W}}^{[1]} )
 = \text{span} (\textbf{G}^{[32]} \underline{\textbf{W}}^{[2]} ).
\end{equation}
Let $\textbf{W}^{[ij]} = [ \textbf{w}_1^{[ij]}~\dots~\textbf{w}_d^{[ij]}  ]$,
$i \in \{1,2,3\}$ and $j \in \{1,\dots,K\}$. We rewrite (\ref{eq7111}) as
\begin{multline}
\text{span} \Big(
\begin{bmatrix}\textbf{H}_1^{[21]}~
\dots~ \textbf{H}_1^{[2K]}
\end{bmatrix}
\begin{bmatrix}
\textbf{w}_k^{[21]}\\ \dots\\ \textbf{w}_k^{[2K]}
\end{bmatrix}
\Big) \\= \text{span} \Big(
\begin{bmatrix}\textbf{H}_1^{[31]}~
\dots~ \textbf{H}_1^{[3K]}
\end{bmatrix}
\begin{bmatrix}
\textbf{w}_k^{[31]}\\ \dots\\ \textbf{w}_k^{[3K]}
\end{bmatrix}
\Big),
\end{multline}
where $k\in \{1,\dots,d\}$, i.e.,
\begin{multline}
 \text{span} (  \textbf{H}_1^{[21]} \textbf{w}_k^{[21]} + ~\dots~ +
 \textbf{H}_1^{[2K]} \textbf{w}_k^{[2K]} ) \\ = \text{span}(
  \textbf{H}_1^{[31]} \textbf{w}_k^{[31]}  + ~\dots~ + \textbf{H}_1^{[3K]} \textbf{w}_k^{[3K]}
  ).
\end{multline}
So, there exist non-zero $\alpha_{1k}$ and $\alpha_{2k}$ satisfy
\begin{multline}
 \alpha_{1k} \Big(  \textbf{H}_1^{[21]} \textbf{w}_k^{[21]} +~\dots~ +
 \textbf{H}_1^{[2K]} \textbf{w}_k^{[2K]} \big) \\= \alpha_{2k}\big(
  \textbf{H}_1^{[31]} \textbf{w}_k^{[31]}  + ~\dots~ +  \textbf{H}_1^{[3K]} \textbf{w}_k^{[3K]}
  \Big).
\end{multline}
i.e.,
\begin{multline}
\text{dim}\Big\{ \text{span}\big( \big[ \textbf{H}_1^{[21]} \textbf{w}_k^{[21]}
~~\dots~~  \textbf{H}_1^{[2K]} \textbf{w}_k^{[2K]}~~\\
 \textbf{H}_1^{[31]} \textbf{w}_k^{[31]} ~~\dots~~
\textbf{H}_1^{[3K]} \textbf{w}_k^{[3K]} \big] \big) \Big\}=2K-1.
\end{multline}
Then we have
\begin{multline}\label{eq84}
\text{dim}\Big\{ \text{span}\big( \big[ \textbf{H}_1^{[21]} \textbf{W}^{[21]}
~~\dots~~  \textbf{H}_1^{[2K]} \textbf{W}^{[2K]} ~~\\
~~~~~~~~~~~~~~~~~~~~~~~~~~~~~~ \textbf{H}_1^{[31]} \textbf{W}^{[31]} ~~\dots~~
\textbf{H}_1^{[3K]} \textbf{W}^{[3K]} \big] \big) \Big\}\\
= \text{dim}\Big\{ \text{span}\big( \big[ \textbf{H}_1^{[21]}
\textbf{w}_1^{[21]} ~~\dots~~  \textbf{H}_1^{[2K]} \textbf{w}_1^{[2K]}~~~~~~~~~~~~~~~~~~~~~~~\\
~~~~~~~~~~~~~~~~~~~~~~~~~~~~~~ \textbf{H}_1^{[31]} \textbf{w}_1^{[31]}
~~\dots~~
\textbf{H}_1^{[3K]} \textbf{w}_1^{[3K]} \big] \big) \Big\}\\
+\dots~~~~~~~~~~~~~~~~~~~~~~~~~~~~~~~~~~~~~~~~~~~~~~~~~~~~~~~~~~~~~~~~~~~\\
+ \text{dim}\Big\{ \text{span}\big( \big[ \textbf{H}_1^{[21]}
\textbf{w}_d^{[21]} ~~\dots~~ \textbf{H}_1^{[2K]} \textbf{w}_d^{[2K]}~~~~~~~~~~~~~~~~~~~~~~~~~~~\\
~~~~~~~~~~~~~~~~~~~~~~~~~~~~ \textbf{H}_1^{[31]} \textbf{w}_d^{[31]} ~~\dots~~
\textbf{H}_1^{[3K]} \textbf{w}_d^{[3K]} \big] \big) \Big\}\\
=(2K-1)d.~~~~~~~~~~~~~~~~~~~~~~~~~~~~~~~~~~~~~~~~~~~~~~~~~~~~~~~~~~~~~~~~~~
\end{multline}
So, the interference space at BS1 can be aligned into a vector space of
$(2K-1)d$ dimensions. Along the same way, we also have
\begin{multline}\label{eq85}
\text{dim}\Big\{\text{span}\big( \big[
\textbf{H}_2^{[11]}\textbf{W}^{[11]}~~\dots~~
\textbf{H}_2^{[1K]}\textbf{W}^{[1K]}~~ \textbf{H}_2^{[31]}\textbf{W}^{[31]}~~\\
\dots~~ \textbf{H}_2^{[3K]}\textbf{W}^{[3K]} \big] \big) \Big\} = (2K-1)d,
\end{multline}
\begin{multline}\label{eq86}
\text{dim}\Big\{\text{span}\big( \big[
\textbf{H}_3^{[11]}\textbf{W}^{[11]}~~\dots~~
\textbf{H}_3^{[1K]}\textbf{W}^{[1K]}~~ \textbf{H}_3^{[21]}\textbf{W}^{[21]}~~\\
\dots~~ \textbf{H}_3^{[2K]}\textbf{W}^{[2K]} \big] \big) \Big\} = (2K-1)d.
\end{multline}
Then, each BS needs a vector space of $M \geq [(2K-1) + K]d$ dimensions to
decode $Kd$ streams sent from its $K$ cell-edge users, i.e., $d \leq \lfloor
\frac{M}{3K-1} \rfloor$ should be satisfied when $M$ and $K$ are given. Then
let $\textbf{V}^{[i]}$, $i \in \{1,2,3\}$, be a subset of (or equal to) the
null space of the interference at each BS, which is a $M \times Kd$ matrix. IUI
elimination matrix can be calculated accordingly. So, if $M =KN$ and $d \leq
\lfloor \frac{M}{3K-1} \rfloor$, totally $3Kd$ DoF is achievable.

\setcounter{equation}{32}

When $M<KN$ and $d \leq \min \big\{\lfloor \frac{M}{3K-1} \rfloor, 3(KN-M)
\big\}$, We combine all the interference signals into a matrix
$\overline{\textbf{H}}$ which is defined as (\ref{eq77}). Let
\begin{multline}
\overline{\textbf{W}} = \Big[ \textbf{W}^{[11]\dag}~ \dots~
\textbf{W}^{[1K]\dag}~ \textbf{W}^{[21]\dag}~ \dots~\\ \textbf{W}^{[2K]\dag}~
\textbf{W}^{[31]\dag}~ \dots~ \textbf{W}^{[3K]\dag} \Big]^{\dag},
\end{multline}
and let
\begin{equation} \label{eq237}
\overline{\textbf{W} } \subseteq \text{null} ( {\overline{\textbf{H}}} ).
\end{equation}
The dimension of ${\overline{\textbf{H}}}$ is $3M \times 3KN$, and the
dimension of the null space of ${\overline{\textbf{H}}}$ is $3(KN-M)$. Let $d
\leq 3(KN-M)$, then there exists a $3KN \times d$ $\overline{\textbf{W} }$
satisfies (\ref{eq237}). Then (\ref{eq84}) - (\ref{eq86}) are also satisfied.
By using the same argument as in the motivating example, $\textbf{W}^{[ij]}$
will be full rank with probability 1.

The dimension of the interference space at each BS is decreased to $(2K-1)d$,
then each BS needs a vector space of $M \geq [(2K-1) + K]d$ dimensions to
decode $Kd$ streams sent from its $K$ cell-edge users, i.e., $d \leq \lfloor
\frac{M}{3K-1} \rfloor$ should be satisfied when $M$ and $K$ are given. ICI
elimination matrix $\textbf{V}^{[i]}$ and IUI elimination matrix
$\textbf{P}^{[i]}$, $i \in \{1,2,3\}$, can be calculated accordingly. Then
totally $3Kd$ streams can be sent simultaneously, i.e., when $M < KN$ and $d
\leq \min \big\{\lfloor \frac{M}{3K-1} \rfloor, 3(KN-M) \big\}$, $3Kd$ DoF is
achievable. For example, when $M=8$, $N=4$, and $K=3$, one stream can be sent
from each user simultaneously. Then $\eta  =9 $ is achievable, while if
orthogonal schemes are used, at most 8 interference-free streams can be sent
simultaneously.
\end{proof}

\section{Numerical Results}
\begin{figure}[!tr]
\centering
\includegraphics[width=9.0cm]{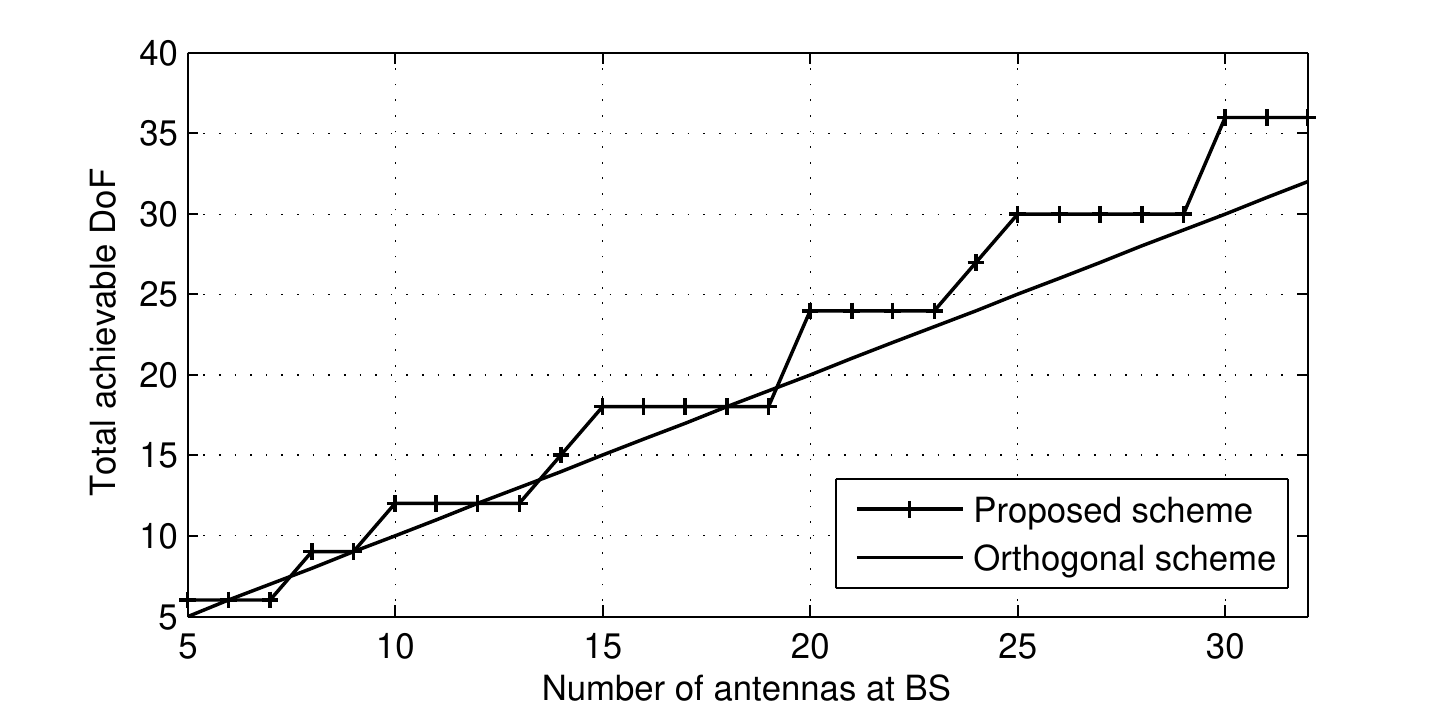}
\caption{The comparison of DoF achievable between our IA based scheme and
orthogonal schemes when $N<M \leq KN$.} \label{fig33}
\end{figure}

The comparison of DoF achievable between our scheme and orthogonal schemes is
presented in Fig.~\ref{fig33} when $N<M \leq KN$ and
\begin{equation*}
K = \arg\max_{K \in \mathbb{R}, N<M} ~3K \cdot \min \big\{\lfloor
\frac{M}{3K-1} \rfloor, 3(KN-M) \big\},
\end{equation*}
where $K$ varies from $2$ to $5$ when $M \leq 32$. It is shown that our scheme
can achieve more DoF compared with orthogonal schemes in most cases. However,
achievable DoF is less than orthogonal schemes when $M =7$, 13, or 19, as some
dimensions are wasted at BSs. If symbol extensions are allowed, even with
constant channel, it is expected that more DoF can be achieved than that of
orthogonal schemes by using similar scheme in \cite{ref8}, and we leave it for
future work.

\ifCLASSOPTIONcaptionsoff
  \newpage
\fi


\begin{thebibliography}{100}

\bibitem{ref1}
V.~R.~Cadambe and S.~A.~Jafar, ``Interference alignment and the degrees of
freedom for the $K$-user interference channel," \emph{IEEE Trans. Inf. Theory},
vol. 54, no. 8, pp. 3425-3441, Aug. 2008.

\bibitem{ref2}
R.~Tresch, M.~Guilland, and E.~Riegler, ``On the achievability of interference
alignment in the $K$-user constant MIMO interference channel," in \emph{Proc.
IEEE Workshop Stat. Signal Process}, pp. 277-280, Cardiff, U.K., Sep. 2009.


\bibitem{ref3}
M.~Razaviyayn, G.~Lyubeznik, and Z.-Q.~Luo, ``On the degrees of freedom
achievalbe through interference alignment in a MIMO interference channel,"
\emph{IEEE Trans. Signal Process.}, vol. 60, no.2, pp. 812-821, Feb. 2012.

\bibitem{ref4}
G. Bresler, D. Cartwright, and D. Tse, ``Settling the feasibility of
interference alignment for the MIMO interference channel: the symmetric case,"
in arXiv:1104.0888v1, Apr. 2011.

\bibitem{ref5}
K. Gomadam, V. R. Cadambe, and S. A. Jafar, ``Approaching the capacity of
wireless networks through distrubted interference alignment," in \emph{Proc. of
IEEE GLOBECOM}, Dec. 2008.


\bibitem{ref6}
C.~Suh, M.~Ho, and D.~Tse, ``Downlink interference alignment," \emph{IEEE
Trans. Commun.}, vol. 59, no. 9, pp. 2616-2626, Sep. 2011.


\bibitem{ref7}
W.~Shin, N.~Lee, J.-B.~Lim, C.~Shin, and K.~Jang, ``On the design of
interference alignment scheme for two-cell MIMO interfering broadcast
channels," \emph{IEEE Trans. on Wireless Communications}, Vol. 10, no. 2, pp.
437-442, Feb. 2011.

\bibitem{ref8}
P. Mohapatra, K.~E.~Nissar, and C.~R.~Murthy, ``Interference alignment
algorithms for the $K$ user constant MIMO interference channel," \emph{IEEE
Trans. Signal Process.}, vol. 59, no. 11, pp. 5499-5508, Nov. 2011.

\end{thebibliography}
\end{document}